\documentclass{ws-ijmpd}
\usepackage{amsmath}
\usepackage{bm}
\begin{document}

\markboth{Bahram Mashhoon}
{Beyond Gravitoelectromagnetism: Critical Speed in Gravitational Motion}

\catchline{}{}{}{}{}

\title{Beyond Gravitoelectromagnetism:\\ Critical Speed in Gravitational Motion}

\author{Bahram Mashhoon}

\address{Department of Physics and Astronomy, University of Missouri-Columbia,\\ Columbia,
Missouri 65211, USA\\
mashhoonb@missouri.edu}

\maketitle

\begin{history}
\received{Day Month Year}
\revised{Day Month Year}
\comby{Managing Editor}
\end{history}

\begin{abstract}
A null ray approaching a distant astronomical source appears to slow down, while a massive particle
speeds up in accordance with Newtonian gravitation. The integration of these apparently
incompatible aspects of motion in general relativity is due to   the existence of a
critical speed. Dynamics of particles moving faster than the critical speed could then be
contrary to Newtonian expectations. Working within the framework of
gravitoelectromagnetism, the implications of the existence of a critical speed are
explored. The results are expected to be significant for high energy astrophysics.
\end{abstract}


\section{Introduction\label{s1}} 

Gravitoelectromagnetism (``GEM") originates from the similarity between Coulomb's law of
electricity and Newton's law of universal gravitation. In the course of the development
of his dynamical theory of the electromagnetic field, Maxwell considered a similar
theory for the gravitational field\cite{1}. He noted that gravitational charges are all
of the same kind and lead to attraction instead of repulsion. On the basis of certain
energy arguments, he then concluded that one could not arrive at a fundamental theory of
gravitation in this way\cite{1}. Several years later, difficulties with the excess
perihelion motion of Mercury led to the phenomenological introduction of a
gravitomagentic force due to mass current in the Sun\cite{2,3}. However, in 1915
Einstein's general relativity provided a beautiful explanation of Mercury's
motion\cite{4}. This involved a relativistic correction to the Newtonian gravitoelectric
potential of the Sun. Soon afterwards, the gravitational influence of the rotation of
the Sun on planetary orbits was determined within general relativity by de
Sitter\cite{5}. This equatorial gravitomagnetic precession turned out to be much smaller
and in the opposite sense as the Einstein gravitoelectric precession\cite{5}.

A field theory that successfully combines Newtonian gravitation with Lorentz invariance
should necessarily contain a gravitomagnetic field\cite{6}. The first general
investigation of the gravitomagnetic field within the framework of general relativity is
due to Thirring\cite{7}. Moreover, the general rate of precession of the orbit of a test
particle in the field of a rotating mass was determined by Lense and Thirring\cite{8}.
An English translation of the Thirring-Lense papers together with a critical commentary
on their results has been provided in Ref.~\refcite{9}. Extensive reviews and discussions
of the various aspects of GEM are contained in Refs.~\refcite{10} --\refcite{12}. On the
observational
side, there has been progress, especially since NASA's GP-B was launched on April 20,
2004.\cite{13}\cdash\cite{15}

The GEM framework will be employed in this paper in order to illustrate the existence of
a critical speed in the dynamics of a test particle moving in a given GEM field. There
are two principal approaches to GEM. The linear perturbation approach to GEM is briefly
presented in Section~\ref{s2} and the general equation of motion of a test particle is
investigated in Section~\ref{s3}. It is shown that this equation of motion contains a
critical speed given by $v_c=c/\sqrt{3}$ in the case of one-dimensional motion.
Section~\ref{s4} is devoted to a more basic invariant treatment of the critical speed;
for this purpose, the spacetime curvature approach to GEM is employed using the
quasi-inertial Fermi coordinates. In this invariant analysis of relative motion, the
critical speed turns out to be $V_c=c/\sqrt{2}$. For relative (``ultrarelativistic")
motion that is faster than this critical speed, the results of recent investigations of
tidal acceleration/deceleration phenomena in the field of a neutron star or a black hole
are briefly summarized. The case of ultrarelativistic fluid flow is very similar and the
relevant equations are derived in \ref{aA}. Finally, Section~\ref{s5} contains
general remarks regarding the gravitational acceleration/deceleration of
ultrarelativistic particles.

\section{Motion in a GEM Field\label{s2}}

Gravitoelectromagnetism provides a useful method for the description of the
gravitational field generated by a slowly rotating ``nonrelativistic" astronomical
source in the linear approximation of general relativity\cite{4}. The spacetime
metric can be written as $g_{\mu\nu}=\eta_{\mu\nu}+h_{\mu\nu}$, where $\eta_{\mu\nu}$ is
the Minkowski metric tensor with signature $+2$ and $h_{\mu\nu}$ is a first-order
perturbation. The gravitational potentials are in general gauge-dependent, i.e.
$h_{\mu\nu}\to h_{\mu\nu }+\epsilon _{\mu,\nu}+\epsilon_{\nu,\mu}$, where $\epsilon_\mu$
is a vector associated with the choice of the background inertial coordinates
$x^\mu=(ct,\mathbf{x}$). In terms of the trace-reversed potentials
$\bar{h}_{\mu\nu}=h_{\mu\nu}-\frac{1}{2}\eta_{\mu\nu}h$, where $h=\text{tr} (h_{\mu\nu})$, the
gravitational field equations can be written as
\begin{equation}\label{eq1} \Box\bar{h}_{\mu\nu}=-\frac{16\pi
G}{c^4}T_{\mu\nu}\end{equation}
after the (``Lorentz") gauge condition $\bar{h}^{\mu\nu}_{\;\;\;\;,\nu}=0$ has been
imposed.

The general retarded solution of \eqref{eq1} involves the particular solution
\begin{equation}
\label{eq2} \bar{h}_{\mu\nu} =\frac{4G}{c^4} \int \frac{T_{\mu\nu}(ct-|\mathbf{x}-\mathbf{x}'|,\mathbf{x}')}{|\mathbf{x}-\mathbf{x}'|}
d^3x'\end{equation}
plus a solution of the homogeneous wave equation that we simply ignore in the present
analysis. For the sources under consideration, $\bar{h}_{00}=4\Phi /c^2$, $\bar{h}_{0i}=-2
A_i/c^2$ and $\bar{h}_{ij}=O(c^{-4})$. Here $\Phi (t,\mathbf{x})$ is the gravitoelectric
potential and $\mathbf{A}(t,\mathbf{x})$ is the gravitomagnetic vector potential. All
terms of $O(c^{-4})$ will be neglected in the metric tensor. Thus
the important quantities of interest in equation \eqref{eq2} are $T^{00}=\rho c^2$ and
$T^{0i}=cj^i$, where $j^\mu =(c\rho , \mathbf{j})$ is the mass-energy current. The
conservation of the mass-energy of the source is assured through the imposition of the
gauge condition, i.e. $j^\mu_{\;\; ,\mu}=0$ follows from
\begin{equation}\label{eq3} \frac{1}{c}\frac{\partial \Phi}{\partial t}+\nabla \cdot
\left(\frac{1}{2}\mathbf{A}\right) =0.\end{equation}
The spacetime metric is thus given by 
\begin{equation}\label{eq4} -ds^2=-c^2\left( 1-2\frac{\Phi }{c^2}\right)
dt^2-\frac{4}{c}(\mathbf{A}\cdot d\mathbf{x})dt+\left(1+2\frac{\Phi}{c^2}\right) \delta_{ij}dx^idx^j.\end{equation}

It is possible to define the gravitoelectric and gravitomagnetic fields in close analogy
with electrodynamics
\begin{equation}\label{eq5} \mathbf{E}=-\nabla \Phi -\frac{1}{c}\frac{\partial}{\partial
t} \left( \frac{1}{2}\mathbf{A}\right), \quad \mathbf{B}=\nabla \times \mathbf{A}.
\end{equation}
It follows from equations \eqref{eq3} and \eqref{eq5} that
\begin{equation}\label{eq6} \nabla \times
\mathbf{E}=-\frac{1}{c}\frac{\partial}{\partial t} \left( \frac{1}{2} \mathbf{B}\right)
,\quad \nabla \cdot \left( \frac{1}{2}\mathbf{B}\right)=0,\end{equation}
while the gravitational field equations \eqref{eq1} reduce to
\begin{equation}\label{eq7} \nabla \cdot \mathbf{E}=4\pi G\rho ,\quad \nabla \times
\left( \frac{1}{2}\mathbf{B}\right)=\frac{1}{c}\frac{\partial }{\partial
t}\mathbf{E}+\frac{4\pi G}{c}\mathbf{j}.\end{equation}
These are the Maxwell equations for the GEM field; they are based on a certain convention that is explained in the following paragraph.

To preserve the electromagnetic analogy as much as possible, it would be convenient to
be able to employ in the GEM case the standard results of electrodynamics using a
special convention. To this end, we assume that the source has gravitoelectric charge
$Q_E=GM$ and gravitomagnetic charge $Q_B=2GM$, where $M$ is the total mass of the
source. It follows that the gravitomagnetic dipole moment of the source is
$G\mathbf{J}/c$, where $\mathbf{J}$ is the total angular momentum of the source. Thus far
from the source $(r=|\mathbf{x}|\gg GM/c^2)$
\begin{equation}\label{eq8} \Phi \sim \frac{GM}{r},\quad \mathbf{A}\sim
\frac{G}{c}\frac{\mathbf{J}\times \mathbf{x}}{r^3},\end{equation} which are consistent
with equations \eqref{eq2} and \eqref{eq3}. Moreover, a test particle of inertial mass
$m$ has gravitoelectric charge $q_E=-m$ and gravitomagnetic charge $q_B=-2m$ in this
convention. We note that the signs of $(q_E,q_B)$ are opposite to those of $(Q_E,Q_B)$
to preserve the attractive nature of gravity; furthermore, the ratio of gravitomagnetic
charge to the gravitoelectric charge is always $2$, since the linear approximation of
general relativity involves a spin-2 field. This is consistent with the fact that the
ratio of the magnetic charge to the electric charge of a particle is unity in Maxwell's
spin-1 electrodynamics\cite{16}. The magnetic charge employed here should be distinguished
from the magnetic monopole strength, which is strictly zero throughout this paper.

The geodesic equation of motion of a test particle in the stationary GEM field of a
source with $\partial \Phi /\partial t=0$ and $\partial \mathbf{A} /\partial t=0$ may be
written in  a form analogous to the Lorentz force law, 
\begin{equation}\label{eq9}
m\frac{d\mathbf{v}}{dt}=-m\mathbf{E}-2m\frac{\mathbf{v}}{c}\times
\mathbf{B},\end{equation}
when velocity-dependent terms of order higher than $v/c$ are neglected. To go beyond the
GEM analogy, let us explore the significance of the terms that have been neglected in
equation~\eqref{eq9}. The geodesic equation for a particle with proper time $\tau$ and four-velocity $u^\mu
=dx^\mu /d\tau =\gamma (1,\mathbf{v}/c)$ is given by
\begin{equation}\label{eq10} \frac{du^\mu}{d\tau} + \Gamma ^\mu_{\rho \sigma} u^\rho
u^\sigma =0.\end{equation}
Once the linear GEM field is given, it is possible to employ equation~\eqref{eq10} for
the motion of particles of any speed in this field; for instance, one may consider the
force-free motion of cosmic rays in the gravitational field of the Earth. For the sake
of simplicity, we choose units such that $c=1$ in what follows.  For the GEM field, the
Christoffel symbols are
\begin{align}\label{eq11} \Gamma ^0_{0\mu} &=-\Phi_{,\mu},\quad \Gamma
^0_{ij}=2A_{(i,j)}+\delta_{ij}\Phi_{,0},\\
\label{eq12} \Gamma^i_{00}&=-\Phi_{,i}-2A_{i,0},\quad
\Gamma^i_{0j}=\delta_{ij}\Phi_{,0}+\epsilon_{ijk}B^k,\\
\label{eq13}
\Gamma^i_{jk}&=\delta_{ij}\Phi_{,k}+\delta_{ik}\Phi_{,j}-\delta_{jk}\Phi_{,i}.\end{align}
The physical consequences of the geodesic equation~\eqref{eq10} with the connection
given by equations~\eqref{eq11}-\eqref{eq13} are explored in the following section.

\section{Critical Speed\label{s3}}

It is straightforward to show that the components of equation~\eqref{eq10} can be
expressed as
\begin{align}\label{eq14}\frac{1}{\gamma}\frac{d\gamma}{dt}&=(1-v^2) \Phi_{,0}+2v^i[\Phi_{,i}-A_{(i,j)}v^j],\\
\frac{dv^i}{dt}&= (1+v^2)\Phi_{,i}-2(\mathbf{v}\times
\mathbf{B})_i +2A_{i,0}-v^i(3-v^2)\Phi_{,0}\notag\\
&\quad +2 v^iv^j[A_{(j,k)}v^k-2\Phi_{,j}].\label{eq15}
\end{align}
Here we have used the fact that $du^\mu/d\tau =\gamma du^\mu/dt$ and then separated the equations for $\gamma$ and $\mathbf{v}$.
The worldline of the test particle is timelike; therefore, it follows from $u^\mu u_\mu
=-1$ that
\begin{equation}\label{eq16} \frac{1}{\gamma^2}=1-v^2-2(1+v^2)\Phi +4\mathbf{v}\cdot
\mathbf{A}.\end{equation}

Consider a stationary axisymmetric source such that for $r\gg GM/c^2$ the potentials are
given by equation~\eqref{eq8}. In this case, equation~\eqref{eq15} may be expressed as
\begin{align}
\frac{d\mathbf{v}}{dt}&=-\frac{GM}{r^3}[(1+v^2)\mathbf{x}-4 (\mathbf{x}\cdot
\mathbf{v})\mathbf{v}]-\frac{2G}{r^5} [r^2\mathbf{J}\times \mathbf{v}\notag\\
&\quad -3(\mathbf{J}\cdot \mathbf{x})\mathbf{L}+3(\mathbf{J}\cdot
\mathbf{L})(\mathbf{x}\cdot \mathbf{v})\mathbf{v}],\label{eq17}\end{align}
where $\mathbf{L}=\mathbf{x}\times \mathbf{v}$. For one-dimensional motion along the
rotation axis of the source, equation~\eqref{eq17} reduces to 
\begin{equation}\label{eq18} \frac{d\mathbf{v}}{dt}=-\frac{GM\mathbf{x}}{r^3}(1-3v^2).\end{equation}
This equation contains a critical speed $v_c=1/\sqrt{3}$; that is, for motion with
$v<v_c$, we have the standard attractive force of gravity familiar from Newtonian physics,
while for $v=v_c$, the particle experiences no force and for $v>v_c$ the gravitational
attraction turns to repulsion. These results are valid in the {\it linear} approximation
for the gravitational field under consideration here. It is interesting to note that if
we use instead the standard post-Newtonian approximation scheme for the field of the
source, the factor $(1-3v^2)$ in equation~\eqref{eq18} becomes $(1-3v^2-4\Phi)$, where
$\Phi =GM/r\ll 1$.\cite{17}

Equation~\eqref{eq18} has the exact solution $(v=dr/dt)$
\begin{equation}\label{eq19} v^2=v_c^2-(v^2_c-v^2_\infty )e^{-6\Phi};\end{equation}
however, in the linear approximation, $e^{-6\Phi}\approx 1-6\Phi$, and
equation~\eqref{eq19} is valid only in the form
\begin{equation}\label{eq20}v^2=v^2_\infty +\frac{2GM}{r}(1-3v^2_\infty),\end{equation}
so that the speed of an infalling test particle increases, remains constant or decreases
depending on whether $v_\infty$ is less than, equal to or more than $v_c=1/\sqrt{3}$,
respectively. In the nonrelativistic limit, $1-3v^2_\infty\to 1$ and
equation~\eqref{eq20} reduces to the standard result of Newtonian gravitation.

It is crucial to recognize here that highly relativistic outflows in the form of
astrophysical jets are expected to occur along the rotation axis of the central massive source. Thus an
equation such as \eqref{eq18}, though valid for motion along the rotation axis, could
nevertheless play a rather significant role in the gravitational physics of phenomena
associated with jets.

In connection with jets, it is useful to consider radial motion in the exterior Kerr
spacetime. The Kerr metric is given by
\begin{align} -ds^2&=-dt^2+\Sigma
\left(\frac{1}{\Delta}d\hat{\rho}^2+d\vartheta^2\right)+(\hat{\rho}^2+a^2)\sin^2\vartheta d\phi^2\notag\\
&\quad + 2GM\frac{\hat{\rho}}{\Sigma }(dt-a\sin^2\vartheta
d\phi)^2,\label{eq21}\end{align}
where $\Sigma =\hat{\rho}^2+a^2\cos^2\vartheta$ and $\Delta
=\hat{\rho}^2-2GM\hat{\rho}+a^2$ in Boyer-Lindquist coordinates $(t,\hat{\rho},\vartheta
,\phi)$. Here $M$ is the mass and $a=J/M$ is the specific angular momentum of the
source. The geodesic equations of motion for a test particle moving along the rotation
axis reduce to
\begin{equation}\label{eq22}
\frac{dt}{d\tau}=\hat{\gamma}\frac{\hat{\rho}^2+a^2}{\hat{\rho}^2-2GM
\hat{\rho}+a^2},\quad \frac{d\hat{\rho}}{d\tau}=\pm \sqrt{\hat{\gamma}^2-1+\frac{2GM
\hat{\rho}}{\hat{\rho}^2 +a^2}},\end{equation}
where $\hat{\gamma}$ is an integration constant. When $\hat{\gamma}\geq 1$, it is the
Lorentz factor of the test particle as measured by the static inertial observers at
spatial infinity, i.e. $\hat{\gamma }^{-2}=1-v^2_\infty$. It follows from
system~\eqref{eq22} that
\begin{equation}\label{eq23} \left( \frac{d\hat{\rho}}{dt}\right)^2=v^2_\infty \left(
1+12\frac{G^2M^2}{\hat{\rho}^2}\right) +\frac{2GM}{\hat{\rho}}\left( 1-3v^2_\infty
-\frac{4GM}{\hat{\rho}}\right)+O\left( \frac{1}{\hat{\rho}^3}\right).\end{equation}
To compare with our GEM results, this equation must be expressed in terms of the
isotropic radial coordinate $r$,
\begin{equation}\label{eq24} \hat{\rho}=r\left( 1+\frac{GM}{2r}\right)^2.\end{equation}
Thus equation~\eqref{eq20} can be recovered, in terms of either $\hat{\rho}$ or $r$, to
linear order in the gravitoelectric potential.

Shapiro's radar echo delay experiments have demonstrated that photons in effect slow
down in the gravitational field of a mass $M$. In Newtonian gravity, however, massive
test particles speed up as they approach a gravitational source. It turns out that this
is also the case in general relativity if the speed of the particle at infinity is below
a critical speed. Otherwise, as $v_\infty$ approaches the speed of light the particle
has lightlike behavior. That a particle with an initial speed $v_\infty$ above a
critical speed slows down in the Schwarzschild field as the particle radially approaches
the source was first demonstrated in Refs.~18 and 19. The critical
speed $v_c=1/\sqrt{3}$ was discussed by Carmeli\cite{18}, who derived
equation~\eqref{eq20} for radial motion in the exterior Schwarzschild field. 
The propulsion aspects of this result have recently received attention.\cite{20}
Using a more invariant approach, Jaffe and Shapiro\cite{19} discussed a transition
velocity, $v_t\simeq 1/\sqrt{2}$, in connection with the radial motion of particles in
the Schwarzschild field. They argued that the coordinate speed should be replaced with
$dl/dt$, where $dl^2=\gamma_{ij}dx^idx^j$ and $\gamma_{ij}=g_{ij}-g_{0i}g_{0j}/g_{00}$.
This results, working to linear order in the Schwarzschild spacetime, in an equation of
the same form as equation~\eqref{eq20} except that the factor $(1-3v_\infty^2)$ is
replaced by $(1-2v^2_\infty)$.

It is clear that by employing physically admissible coordinate systems one can obtain
different numerical values for the critical speed. The critical speeds $1/\sqrt{3}$ and
$1/\sqrt{2}$ have also been discussed in Ref.~\refcite{21}. Nevertheless, it is important
to remark here that with respect to the basic class of static (generally noninertial)
observers in the exterior Schwarzschild spacetime, the local speed of an infalling
particle monotonically increases toward unity, while the local speed of light is always
equal to unity.

In the standard post-Newtonian approximation scheme, one is limited to the slow-motion
weak-field regime and therefore the coordinate quantities in the equations of motion are
assumed to be physically significant as they are observable in the standard
interpretation. Here, however, we have been using the linear post-Newtonian scheme for
the determination of the field generated by the central source, while the test particles
can travel at any speed. Thus to study further the predictions of general relativity in
this case, we must construct appropriate invariants that would represent actually
measurable quantities. To this end, we find it convenient to study {\it relative} motion
in the quasi-inertial Fermi coordinate system\cite{22} established about the motion
of a reference observer. This manifestly invariant approach is described in the next
section.

\section{Fermi Coordinates\label{s4}}

Consider an observer $\mathcal{O}$ following a geodesic in the gravitational field of an
astronomical source. Let $\lambda^\mu_{\;\;(\alpha)}$ be the observer's local
orthonormal tetrad frame that is parallel propagated along the worldline of
$\mathcal{O}$. That is, $\lambda^\mu_{\;\;(0)}=dx^\mu /d\tau$ is the observer's
four-velocity vector as well as its local temporal axis and $\lambda^\mu _{\;\;(i)}$,
$i=1,2,3$ are the spatial unit gyro directions that form the local spatial frame of the
observer. Here $\tau$ is the proper time of $\mathcal{O}$. To describe physical phenomena
relative to $\mathcal{O}$, it is natural to construct a Fermi coordinate system along
its worldline. This is a geodesic coordinate system based on the observer's local frame.
Indeed, an event $P$ with Fermi coordinates $X^\mu=(T,\mathbf{X})$ can be orthogonally
connected to the worldline of the reference observer at $P_0$ by a unique spacelike
geodesic of proper length $\sigma$. Let $\tau$ be the proper time along the reference
trajectory at $P_0$ and $\eta^\mu =(dx^\mu /d\sigma )_0$ be the unit tangent vector to
the spacelike geodesic at $P_0$; then, $T=\tau$ and $X_i=\sigma \eta_\mu
\lambda^\mu_{\;\;(i)}$. Thus,  $\mathcal{O}$ has Fermi
coordinates $(\tau ,\boldsymbol{0})$, so that the reference observer always occupies the
spatial origin of the Fermi coordinate system. The spacetime metric in Fermi coordinates
is given by 
\begin{align} {^Fg_{00}} &= -1-\;{^FR_{0i0j}}(T)X^iX^j+\dots ,\label{eq25}\\
{^Fg_{0i}}&=-\frac{2}{3}\;{^FR_{0jik}}(T)X^jX^k+\dots ,\label{eq26}\\
{^Fg_{ij}}&= \delta_{ij}-\frac{1}{3}\;{^FR_{ikjl}}(T)X^kX^l +\dots ,\label{eq27}\end{align}
where
\begin{equation}{^FR_{\alpha \beta \gamma \delta}}(T)=R_{\mu\nu\rho\sigma} \lambda^\mu_{\;\;(\alpha)}
\lambda^\nu_{\;\;(\beta)} \lambda^\rho_{\;\;(\gamma)} \lambda^\sigma _{\;\;(\delta
)}\label{eq28}\end{equation}
is the projection of the Riemann tensor on the local tetrad frame of observer
$\mathcal{O}$. The Fermi coordinates are admissible in a cylindrical region with
$|\mathbf{X}|<\mathcal{R}$ along the reference worldline such that $\mathcal{R}(T)$ is a
certain minimum radius of curvature of spacetime.

The equation of motion of a free test particle in the Fermi coordinate system can be
expressed as
\begin{equation}\label{eq29} \frac{dU^\mu}{ds}+\;{^F\Gamma^\mu_{\alpha\beta}} U^\alpha
U^\beta =0,\end{equation}
where $U^\mu =dX^\mu /ds=\Gamma (1,\mathbf{V})$ is the particle's four-velocity. The
geodesic motion of the particle is always timelike; therefore,
\begin{equation}\label{eq30} \Gamma^{-2}=-\;{^Fg_{00}}-2\;{^Fg_{0i}}
V^i-\;{^Fg_{ij}}V^iV^j>0.\end{equation}
It follows from equation~\eqref{eq29} that
\begin{align} &\frac{1}{\Gamma}\frac{d\Gamma}{dT}=-\;{^F\Gamma^0_{\alpha\beta}}
\frac{dX^\alpha}{dT}\frac{dX^\beta}{dT},\label{eq31}\\
&\frac{d^2X^i}{dT^2}+({^F\Gamma^i_{\alpha\beta}} -\;{^F\Gamma^0_{\alpha\beta}} V^i)
\frac{dX^\alpha}{dT}\frac{dX^\beta}{dT}=0.\label{eq32}\end{align}

The equation of relative motion~\eqref{eq32} can be expressed to linear order in
distance away from the reference observer $\mathcal{O}$ by using the terms given
explicitly in the metric tensor \eqref{eq25}-\eqref{eq27}. The result is
\begin{align}&\frac{d^2X^i}{dT^2}+\;{^FR_{0i0j}}X^j+2\;{^FR_{ikj0}}V^kX^j\notag\\
&\quad +\frac{2}{3}(3\;{^FR_{0kj0}}V^iV^k+\;{^FR_{ikjl}} V^kV^l+\;{^FR_{0kjl}}V^iV^kV^l)X^j=0,\label{eq33}\end{align}
while the modified Lorentz factor is given by
\begin{align}\frac{1}{\Gamma^2}&=1-V^2+\;{^FR_{0i0j}}X^iX^j+\frac{4}{3}\;{^FR_{0jik}}X^jV^iX^k\notag\\
&\quad + \frac{1}{3}\;{^FR_{ikjl}}V^iX^kV^jX^l.\label{eq34}\end{align}
Neglecting the velocity-dependent terms in equation~\eqref{eq33}, one recovers the Jacobi
equation. It is important to note that the treatment of test particle motion can be
generalized to the flow of a perfect fluid as in \ref{aA}.

To interpret the generalized Jacobi equation~\eqref{eq33}, let us first proceed by
analogy with the linear perturbation theory of Section~\ref{s2} and note that the metric
in the Fermi frame reduces to the Minkowski metric along the reference trajectory;
hence, one may write $\;{^Fg_{00}}=-1+2\Phi$ and $\;{^Fg_{0i}}=-2A_i$, where
\begin{equation}\label{eq35} \Phi =-\frac{1}{2}\;{^FR_{0i0j}}X^iX^j,\quad
A_i=\frac{1}{3}\;{^FR_{0jik}}X^jX^k.\end{equation}
Moreover, the corresponding GEM fields are obtained from equation~\eqref{eq5}, so that
to lowest order
\begin{equation}\label{eq36} E_i=\;{^FR_{0i0j}}X^j,\quad
B_i=-\frac{1}{2}\epsilon_{ijk}\;{^FR_{jk0l}}X^l.\end{equation}
It follows that equation~\eqref{eq33} can be written as
\begin{equation}\label{eq37} m\frac{d\mathbf{V}}{dT}=q_E\mathbf{E} +q_B \mathbf{V}\times
\mathbf{B} +O(V^2),\end{equation}
where $q_E=-m$ and $q_B=-2m$ as before. Though this analog of the Lorentz force law is
extremely interesting and leads to the physical interpretation of the Bel and
Bel-Robinson tensors\cite{23}, it is important to recognize that the higher-order
terms in equation~\eqref{eq37} become rather significant for ultrarelativistic motion.
To see this, let us imagine a circumstance where one-dimensional motion---along the $Z$
direction, say---is possible. Then it follows from the symmetries of the Riemann tensor
that equation~\eqref{eq33} reduces to \begin{equation}\label{eq38} \frac{d^2Z}{dT^2}
+k(T)(1-2\dot{Z}^2)Z=0,\end{equation}
where $\dot{Z}=dZ/dT$ and $k(T)=\;{^FR_{TZTZ}}$. The generalized Jacobi
equation~\eqref{eq38} has solutions for $\dot{Z}=\pm 1/\sqrt{2}$ such that the relative
motion within the Fermi frame is uniform. Below the critical speed $V_c=1/\sqrt{2}$, the
relative motion is similar to what is expected from relativistic tides in accordance
with the Jacobi equation. The situation changes drastically, however, for
ultrarelativistic relative motion above the critical speed $V_c$. If $k(T)<0$, then an
initially ultrarelativistic particle decelerates with respect to the reference particle
and asymptotically approaches uniform motion with critical speed $V_c$. If $k(T)>0$,
then an initially ultrarelativistic particle accelerates with respect to the reference
particle.\cite{24} These results for the motion of free particles can be extended to
fluid flow (see \ref{aA}).

Ultrarelativistic particles are expected to be produced in abundance in the central
engines of active galactic nuclei as well as supernova explosions and $X$-ray sources.
It is therefore natural to expect that the tidal acceleration/deceleration of
ultrarelativistic particles could have significant implications for the physics of
astrophysical jets and high-energy cosmic rays.

Assuming that the gravitational field of the central source may be described by the Kerr
field, consider a Fermi coordinate system established along reference escape
trajectories on the axis of rotation given by equation~\eqref{eq22} with
$\hat{\gamma}\gtrsim 1$. Such particles with $\hat{\gamma}$ near unity provide an
ambient medium surrounding the central source. High-energy particles originating near
the poles of the central collapsed configuration and moving outward along the rotation
axis faster than the critical speed $1/\sqrt{2}\approx 0.7$ relative to the ambient
medium experience significant tidal deceleration\cite{25}, while such
ultrarelativistic particles moving outward normal to the rotation axis experience
significant tidal acceleration relative to the ambient medium. In a series of recent
papers, these phenomena have been investigated in some detail.\cite{24}\cdash\cite{27} It
turns
out that tidal deceleration occurs for outflow velocities within a cone of half angle
$\theta \approx \tan^{-1}\sqrt{2}\approx 55^\circ$ around the rotation
axis.\cite{26,27} On the other hand, tidal acceleration occurs for outflow
velocities outside the critical velocity cone.\cite{26,27}

A complete analysis must of course include plasma and radiation effects as well.
Nevertheless, preliminary results appear to be consistent with observations of
microquasars and {\it Chandra} studies of certain neutron star $X$-ray sources in our
Galaxy.\cite{25}\cdash\cite{27}

\section{Discussion\label{s5}}

Using the main approaches to GEM , we have demonstrated the existence of a critical speed
that could have fundamental
consequences for the gravitational physics of ultrarelativistic flows. The physical origin
of the critical speed can be traced back to a ``conflict" between the intuitively
expected behavior of particles based on Newtonian gravitation and the behavior of light
in a gravitational field; indeed, both of these aspects are properly integrated within
general relativity.

Imagine a light ray propagating from event $P_1:(t_1,\mathbf{x}_1)$ to event
$P_2:(t_2,\mathbf{x}_2)$ in a global inertial system of reference; hence,
$t_2-t_1=|\mathbf{x}_2-\mathbf{x}_1|$. If the spacetime is now slightly perturbed by the
presence of a GEM field as in Section~\ref{s2}, then the arrival of the signal at $P_2$
is accompanied mainly by a Shapiro gravitoelectric {\it delay} given by
\begin{equation}\label{eq39} \Delta _{GE}=2\int^{P_2}_{P_1}\Phi\; d\ell ,\end{equation}
where $d\ell$ is the element of length along the path of the unperturbed
ray.\cite{28}
This is intimately related to the fact that the propagation of
electromagnetic waves in the gravitational medium takes place with an
effective index of refraction given by\cite{29} $n\approx   1 + 2\Phi$     for   $\Phi  << 1$. On
the other hand, the kinetic energy of a particle increases as it gets closer to a
gravitational source according to the Newtonian theory of gravitation. The conflict is
resolved in general relativity through the existence of a critical speed such that
``ultrarelativistic" particle motion with speed above this critical speed has
significant features that violate our nonrelativistic intuition based on Newtonian
gravitation. In the linear post-Newtonian approach, the critical coordinate speed is
$1/\sqrt{3}$; however, for relative motion in Fermi coordinates, the invariant critical
speed is $1/\sqrt{2}$. The latter case has been the subject of recent studies due to its direct observational significance.

The most significant consequence of the existence of the critical speed is a possible
gravitational mechanism for the acceleration/deceleration of ultrarelativistic particles.
This mechanism can be elucidated by comparing and contrasting the gravitational and
electromagnetic interactions. The motion of a charged particle in an external
electromagnetic field in Minkowski spacetime is given by
\begin{equation}\label{eq40}\frac{d}{dt}\left( \frac{\mathbf{v}}{\sqrt{1-v^2}}\right)
=\frac{q}{m}(\boldsymbol{\mathcal{E}}+\mathbf{v}\times\boldsymbol{\mathcal{B}})\end{equation}
in accordance with the Lorentz force law. This equation may be written as
\begin{equation}\label{eq41}
\frac{d\mathbf{v}}{dt}=\frac{q}{m}\sqrt{1-v^2}[\boldsymbol{\mathcal{E}}-\mathbf{v}(\mathbf{v}\cdot
\boldsymbol{\mathcal{E}})+\mathbf{v}\times\boldsymbol{\mathcal{B}}],\end{equation}
which indicates that the closer $v$ gets to unity, the more difficult it will be to
change the velocity of the particle. For a finite external electromagnetic field, the
motion of a charged particle with $m\neq 0$ is effectively uniform for $v$ approaching
unity. Thus the critical speed that appears in the electromagentic case is the
fundamental
speed, namely, unity.

The gravitational analog of equation~\eqref{eq40} in the quasi-inertial Fermi coordinate
system is
\begin{equation}\label{eq42}\frac{d}{dT}(\Gamma V^i)=-\Gamma (\;{^F\Gamma
^i_{00}}+2\;{^F\Gamma^i_{0j}}V^j+\;{^F\Gamma^i_{jk}}V^jV^k).\end{equation}
while the analog of equation~\eqref{eq41} is given by equation~\eqref{eq32}. The fact
that the right-hand side of equation~\eqref{eq42} is proportional to the modified
Lorentz factor $\Gamma$ is responsible for the bending of ultrarelativistic particle orbits
by a gravitational field---indeed, this deflection approaches that of light in the
appropriate limit. Similarly, equation~\eqref{eq32} implies that it is in general
possible to change the velocity of ultrarelativistic particles considerably in a
gravitational field, since the analog of $\sqrt{1-v^2}$ in equation~\eqref{eq41}
involving the fundamental speed is missing in equation~\eqref{eq32}; indeed, there is a
critical speed of $1/\sqrt{2}$, but it appears in a different way. An even more striking
illustration of this circumstance
can be given for one-dimensional motion.

Consider a situation where $\boldsymbol{\mathcal{E}}$ and $\boldsymbol{\mathcal{B}}$ are
parallel and
the charged particle moves along the field lines. Then equation~\eqref{eq41} reduces to
\begin{equation}\label{eq43} \frac{d\mathbf{v}}{dt}=\frac{q}{m}
(1-v^2)^{3/2}\boldsymbol{\mathcal{E}},\end{equation}
which may be compared and contrasted with equations~\eqref{eq18} and \eqref{eq38}. While
the ultimate (``critical") speed in equation~\eqref{eq43} severely restricts the
magnitude of acceleration of a particle with $v\to 1$, this is not the case for
gravitational motion in equations~\eqref{eq18} and \eqref{eq38}, which exhibit instead
critical speeds of $1/\sqrt{3}$ and $1/\sqrt{2}$, respectively.

The appearance of the critical factor $( 1 - 2 \dot{Z}^2 )$ in equation \eqref{eq38} is
due to the fact that in the quasi-inertial Fermi coordinates, one finds that
to lowest order
\begin{equation}\label{eq44} {^F\Gamma^i_{00}} =\; {^F\Gamma^0_{0i}}
=-\frac{1}{2}(\;{^Fg_{00}})_{,i}.\end{equation}
Indeed,
the first term in the critical factor originates with    ${^F\Gamma^i_{00}}$,
while the second term originates with   ${^F\Gamma^0_{0i}}$. The factor $2$ in the
latter term is due to the fact that   ${^F\Gamma^0_{0i}}$    occurs twice, since       
${^F\Gamma^0_{0i}}= {^F\Gamma^0_{i0}}$; hence, the
tensorial character of general relativity is responsible for the appearance
of the factor of $2$ in $( 1 - 2 \dot{Z}^2  )$. Thus the circumstance that the critical
speed is unity in electrodynamics while it is $1/\sqrt{2}$ in gravitation is related
to the spin-1 nature of the electromagnetic field and the spin-2 nature of
the linearized general relativity in the Fermi system, respectively.

Recent investigations of the novel gravitational mechanism for the
acceleration/deceleration of ultrarelativistic particles indicate that this mechanism
may play a fundamental role in high-energy astrophysics.\cite{24}\cdash\cite{27}
Furthermore, the tidal motion of {\it charged} particles has also been recently examined in an
interesting astrophysical context in Ref.~\refcite{30}, where the important role of the
critical speed has been independently confirmed.

\appendix

\section{Gravitational Dynamics of Relativistic Flows\label{aA}}

Imagine a perfect fluid with energy density $\rho$ and pressure $p$ in the gravitational
field considered in Section~\ref{s4}. The energy-momentum tensor of the fluid is given by
\begin{equation}\label{eqA1} T^{\mu\nu}=(\rho +p)u^\mu u^\nu +pg^{\mu\nu},\end{equation}
where $u^\mu$ is the unit four-velocity vector of the fluid. The purpose of this
Appendix is to express the dynamical laws $T^{\mu\nu}_{\;\;\;\; ;\nu}=0$ in the Fermi
coordinate system established about the worldline of the geodesic observer
$\mathcal{O}$.

Let us first note that in a global inertial frame of reference, the dynamical laws may
be expressed as
\begin{align}\label{eqA2} \frac{\partial \tilde{\rho}}{\partial t} +\boldsymbol{\nabla}
\cdot (\tilde{\rho}\mathbf{v})&=\frac{\partial p}{\partial t},\\
\label{eqA3}\tilde{\rho}\frac{d\mathbf{v}}{dt}&=-\boldsymbol{\nabla}p -\frac{\partial p}{\partial t} \mathbf{v},\end{align}
where $\tilde{\rho}=\gamma^2(\rho +p)$ and $\gamma =(1-v^2)^{-1/2}$ is the Lorentz
factor of the fluid. Equation~\eqref{eqA2} follows directly from the $\mu=0$ component
of $T^{\mu \nu}_{\;\;\;\; ;\nu}=0$, while the $\mu =i$ component, when combined with
equation~\eqref{eqA2} implies equation~\eqref{eqA3}. Equations~\eqref{eqA2} and
\eqref{eqA3} are the relativistic continuity and Euler equations, respectively, and
assume their familiar form when the pressure has no explicit dependence upon time.

In the quasi-inertial Fermi coordinates, we expect that the dynamical laws would have the
same form as equations~\eqref{eqA2} and \eqref{eqA3} but modified by the presence of
curvature terms. In fact, let the four-velocity be $\Gamma (1,\mathbf{V})$ as before,
where $\Gamma >0$ is given by equation~\eqref{eq30}. Then with $\tilde{\rho}=\Gamma^2(\rho
+p)$, the modified equations of motion become
\begin{align}\label{eqA4} \frac{\partial \tilde{\rho}}{\partial T} + \boldsymbol{\nabla}\cdot (\tilde{\rho}\mathbf{V}) & = -p_{,\alpha}
\;{^Fg^{0\alpha}}-\tilde{\rho} K,\\
\label{eqA5} \tilde{\rho} \left[ \frac{dV^i}{dT}+ (\;{^F\Gamma^i_{\alpha\beta}}-\;{^F\Gamma^0_{\alpha
\beta}}V^i)\frac{dX^\alpha}{dT}\frac{dX^\beta}{dT}\right] &= -p_{,\alpha}\;{^Fg^{i\alpha}}+
(p_{,\alpha}\;{^Fg^{0\alpha}})V^i,\end{align} where 
\begin{equation}
\label{eqA6} K=\left( \;{^F\Gamma^0_{\alpha
\beta}} \frac{dX^\alpha}{dT}+\;{^F\Gamma^\alpha_{\alpha\beta}}\right)\frac{dX^\beta}{dT}.
\end{equation}
To lowest order, $K$ is given by
\begin{equation}
\label{eqA7} K=\frac{1}{3} (8\;{^FR_{0i0j}}-\;{^FR_{ij}})V^iX^j+\frac{2}{3} \;
{^FR_{0ijk}} V^iV^jX^k.\end{equation}
It follows from these results that equations~\eqref{eqA2} and \eqref{eqA3} are recovered
along the reference trajectory at $(T,\mathbf{0})$, as required. Moreover, in the absence
of pressure equation~\eqref{eqA5} is equivalent to equation~\eqref{eq32} with attendant
consequences regarding the existence of the critical speed $V_c=1/\sqrt{2}$. A complete
analysis of equations~\eqref{eqA4} and \eqref{eqA5} is beyond the scope of this work.

\end{document}